\documentstyle[aps,epsf]{revtex}

\tighten
\newcommand{\beq}{\begin{equation}}
\newcommand{\eeq}{\end{equation}}
\newcommand{\ba}{\begin{eqnarray}}
\newcommand{\ea}{\end{eqnarray}}
\newcommand{\nn}{\nonumber}

\newcommand{\bpt}{\bm p_T^{}}
\newcommand{\bkt}{\bm k_T^{}}

\newcommand{\la}{\langle}
\newcommand{\ra}{\rangle}
\newcommand{\amp}[1]{\la #1 \ra}

\newcommand{\slsh}[1]{\mbox{$\not\! #1$}}
\newcommand{\bm}[1]{\bbox{#1}}

\begin{document}
 
\draft
\title{
\begin{flushright}
\begin{minipage}{4 cm}
\small
hep-ph/0004217\\
RIKEN-BNL preprint
\end{minipage}
\end{flushright}
Double transverse spin asymmetries in vector boson production}

\author{Dani\"el Boer}
\address{\mbox{}\\
RIKEN-BNL Research Center\\
Brookhaven National Laboratory, Upton, New York 11973}

\maketitle
\begin{center}\today \end{center}

\begin{abstract}
We investigate a helicity non-flip double transverse spin asymmetry in vector
boson production in hadron-hadron scattering, which was first
considered by Ralston and Soper at the tree level. It does not involve 
transversity 
functions and in principle also arises in $W$-boson production for which we
present the expressions. The asymmetry 
requires observing the transverse momentum of the vector boson, but it is not 
suppressed by explicit inverse powers of a large energy scale. 
However, as we will show, inclusion of Sudakov factors causes suppression of 
the asymmetry, which increases with energy. 
Moreover, the asymmetry is shown to be
approximately proportional to $x_1 g_1(x_1) \, x_2 \overline g_1(x_2)$, 
which gives rise
to additional suppression at small values of the light cone momentum fractions.
This implies that it is negligible for $Z$ or $W$ production 
and is mainly of interest for $\gamma^*$ at low energies. 
We also compare the asymmetry with other types of 
double transverse spin asymmetries and discuss how to disentangle them.    
\end{abstract}


\section{Introduction}

Double transverse spin asymmetries in high energy hadron-hadron collisions
have attracted much theoretical attention (starting with the early
investigations \cite{Ralst-S-79,Baldracchini}), but no experimental studies
have been performed so far. At the BNL Relativistic Heavy Ion Collider (RHIC) 
transversely polarized
protons will be collided for the first time. Therefore it is important to make
an analysis of the double transverse spin asymmetries. The proposed DESY
HERA-$\vec{N}$ experiment also prompts such a study (see e.g.\
\cite{Korotkov}). For the transversity double spin asymmetry in the Drell-Yan
process such studies have been performed in considerable detail \cite{Martin}. 
In this paper we will mainly investigate another type of double transverse 
spin asymmetry, one which does not involve transversity functions. 

In general, helicity non-flip
quark and gluon states in transversely polarized hadron-hadron scattering 
will lead to power suppression [${\cal O}(1/Q^2)$,
where $Q^2$ is the vector boson virtuality]. In the present study 
we will exploit the fact that if the transverse momentum of the produced 
vector boson is observed, this no longer holds true. 
By observing the transverse
momentum certain azimuthal asymmetries can occur in the cross section without
explicit power suppression. 
The possibility of such an unsuppressed helicity non-flip double transverse
spin asymmetry has been noted before in the
literature and a tree level expression for the case of a virtual photon has
been given \cite{Ralst-S-79,Tangerman-Mulders-95a}. Here we investigate
specifically the case of weak vector boson production and the effect of
inclusion of Sudakov form factors.  

The helicity non-flip nature will allow for double transverse
spin asymmetries even in $W$ production (for which the helicity flip
contribution is absent \cite{Bourrely-Soffer}). 
Unfortunately, such transverse momentum dependent azimuthal asymmetries 
will turn out to suffer from suppression due
to Sudakov factors, which increases with energy, as was forseen in Ref.\ 
\cite{Collins-93b}. We will explore this issue quantitatively in detail and 
we will show that for the azimuthal double transverse spin asymmetry of 
interest, the inclusion of Sudakov factors causes suppression by at least an
order of magnitude compared to the tree level result and 
effectively produces a
power behavior of $1/Q^\alpha$, with $\alpha\approx 0.6$. Moreover, the 
asymmetry will be shown to be
approximately proportional to $x_1 g_1(x_1) \, x_2 \overline g_1(x_2)$, 
which gives rise
to additional suppression at small values of the light cone momentum
fractions. 
The conclusion will be that this asymmetry is of interest mainly at lower
energies, i.e.\ for $\gamma^*$ production. This also leaves the option of
studying possible contributions to double 
transverse spin asymmetries in $W$ production from physics beyond the 
standard model.  

The outline of this paper is as follows. In Sec.\ II we will repeat the 
essentials of the 
transversity double spin asymmetry in the Drell-Yan process in order to
contrast it to the helicity non-flip asymmetry (Sec.\ III). We study the
latter asymmetry in the neutral (Sec.\ IIIA) and in the
charged (Sec.\ IIIB) vector boson case. In order to obtain estimates we will
assume Gaussian transverse momentum dependence of the quarks (discussed in 
Sec.\ IIIC).
We will then include Sudakov form factors in the asymmetry (Sec.\ IIID) and 
estimate
its quantitative effects (Sec.\ IIIE). In Sec.\ IIIF we will comment on 
the possible study of physics beyond the standard model via double
transverse spin asymmetries in $W$ production.      
    
\section{Transversity double spin asymmetry}

The main characteristic of the transversity double transverse spin 
asymmetry of vector boson production in hadron-hadron collisions  
is that the gluon distribution does not contribute. 
Hence, at leading order in an expansion in inverse powers of the hard scale(s)
only the quark transversity distribution function 
\cite{Ralst-S-79,Jaffe-Ji-91} 
(denoted as $h_1, \delta q$ or $\Delta_T q$) contributes. This leads to the
well-known expression for the double transverse spin asymmetry in the 
Drell-Yan process:
\beq
A_{TT}= \frac{\sigma(p^{\uparrow} \, {p^{\uparrow}} \to \ell \, \ell' \, X) 
\, {-} \, \sigma(p^{\uparrow} \, {p^{\downarrow}} \to  \ell \, \ell' \, X)}{
\sigma(p^{\uparrow} \, {p^{\uparrow}} \to \ell \, \ell' \, X) 
\, {+} \, \sigma(p^{\uparrow} \, {p^{\downarrow}} \to  \ell \, \ell' \, X)} =  
\frac{\sin^2 \theta \cos 2 \phi_S^\ell}{1 + \cos^2\theta } 
\frac{\sum_{a,\bar a}e_a^2\; h_1^a(x_1) \; 
\overline h{}_1^a(x_2) }{\sum_{a,\bar a}e_a^2\; f_1^a \; \overline f{}_1^a }.
\label{h1h1}
\eeq
We would 
like to note that the sum is over flavors including anti-flavors, otherwise
one should add a term in both numerator and denominator for the exchange 
$\left(x_1 \leftrightarrow x_2 \right)$, since one can use that $h_1^{\bar a}
= \overline h{}_1^a$. 
The above asymmetry comes from the following azimuthal dependence in the cross
section 
\beq
\frac{d\sigma(p^\uparrow \, p^\uparrow 
\to \ell \, \ell' \, X)}{d\Omega dx_1 dx_2}=
\frac{\alpha^2}{3 Q^2}\;\sum_{a,\bar a}e_a^2\;\Big\{ 
       y\,(1-y) \;|\bm S_{1T}^{}|\;|\bm S_{2T}^{}|\;
\cos(\phi_{S_1}^\ell+\phi_{S_2}^\ell)\;
h_1^a(x_1)\overline h{}_1^a(x_2)  
+ \ldots \Big\}.
\label{h1h1bar}
\eeq
The above is expressed in the so-called Collins-Soper frame \cite{CS} 
(for details see e.g.\ \cite{Boer4,Boer-99}):
\ba
\hat t &\equiv & q/Q, \label{tvector}\\
\hat z &\equiv &\frac{x_1}{Q} 
\tilde{P_1}- \frac{x_2}{Q} \tilde{P_2},
\\
\hat h &\equiv & q_T/Q_T = (q-x_1\, P_1 -x_2\, P_2)/Q_T,\label{hvector}
\ea
where $q$ is the vector boson momentum, $P_i$ are the hadron momenta and 
$\tilde{P_i} \equiv P_i-q/(2 x_i)$.
The azimuthal 
angles lie inside the plane orthogonal to $t$ and $z$. In particular, 
$\phi^\ell$ gives the 
orientation of $\hat l_\perp^\mu \equiv \left( g^{\mu \nu}-\hat t^{ 
\mu} \hat t^{\nu } + \hat z^{
\mu} \hat z^{\nu } \right) l_\nu$, the perpendicular part of the lepton
momentum $l$; $\phi_{S_i}^\ell$ is the angle between $\bm{\hat l}_\perp$ and
$\bm S_{iT}^{}$. 
In the cross sections we also 
encounter a dependence on $y=l^-/q^-$, which in the
lepton center of mass frame equals $y=(1 + \cos \theta)/2$, where 
$\theta$ is the angle of $\hat z$ with respect to the momentum of the 
outgoing lepton $l$.

The perturbative corrections to the double 
transverse spin asymmetry, Eq.\ (\ref{h1h1}), 
have been calculated in \cite{Vogelsang-Weber} and using the
assumption that at low energies the transversity distribution function $h_1$
equals the helicity distribution function $g_1$ or by saturating the Soffer
bound, it has been shown in Ref.\
\cite{Martin} that $A_{TT}$ is expected to be of the order of a percent
at RHIC energies. 

In Ref.\ \cite{Bourrely-Soffer} it was discussed that in the process
$p^\uparrow  \, p^\uparrow \to W \, X$ the transversity distribution cannot
contribute and this is a general feature of chiral-odd functions and
charged current exchange processes (see also \cite{Boer-J-M-99}). This means
that only the suppressed contributions from the twist-3 distribution function
$g_T$ and its gluon analogue $\Delta_T g$ \cite{Ji} 
contribute (these are chiral-even functions; they mix under evolution). 
Of course there are contributions of the transversity functions via quark mass 
terms or via production of other particles that can compensate for the
helicity flip, but these are all of higher order in the strong and/or weak
coupling constants (e.g.\ one can think of $p^\uparrow  \, p^\uparrow \to Z \,
X \to W^+ \, W^- \, X$, but this is negligible at RHIC energies). Since
neither quarks nor gluons 
contribute without suppression to the asymmetry $A_{TT}$ in the process 
$p^\uparrow  \, p^\uparrow \to W \, X$, it might make
this asymmetry a good place to look for
contributions from physics beyond the standard model. For instance, scalar or
tensor couplings of the quarks to the $W$ could in principle produce an 
asymmetry. We will return to this issue at the end of this article. First one 
has to investigate and
estimate another standard model mechanism, namely, there is the possibility 
that the quarks (and gluons) are not exactly collinear to the initial
proton, leading to a helicity non-flip asymmetry without explicit suppression
factors of $1/Q$. In other words, if one measures the cross section 
differential in the transverse momentum of the vector boson, 
either in its angle compared to the other particles 
or in its magnitude, the helicity non-flip double transverse spin 
asymmetry {\em can\/} receive contributions at leading order, even for $W$
production. If one averages over this transverse momentum, then the asymmetry
will vanish, but an (inadvertent) incomplete averaging, for instance due to 
imposed cuts, might still have observable consequences, cf.\ for instance Ref.\
\cite{Mirkes}. Even though we will show that for $W$ production this will not
be a problem since the asymmetry turns out to be negligible, 
for $\gamma^*$ at
lower energies this is important to take into consideration.

\section{Helicity non-flip double transverse spin asymmetry}

If one can measure the cross section differential
in the transverse momentum of the vector boson, 
either in its angle compared to the other particles 
or in its magnitude, then 
there is the possibility to have a double transverse spin 
asymmetry at leading order, in principle even for $W$ production. 
To illustrate this we will make use of the 
formalism pioneered by Ralston and
Soper \cite{Ralst-S-79}, which will be applicable in the region where 
the observed
transverse momentum is small compared to the hard scale(s). Since this is a
tree level formalism, we will later on include the effects of resummation of 
soft gluons by combining it with the approach of Ref.\ \cite{CSS85}.
We focus on the Drell-Yan process, first on neutral vector boson production
and later on charged vector boson production. However, the expressions also
apply to $p^\uparrow \, p^\uparrow \to 2 \, \, 
\text{jets}$ (same formulas with
minor trivial replacements) and a similar analysis can be applied to 
$p^\uparrow
\, p^\uparrow \to \pi \, (\Lambda, \ldots) X$, where the observed hadron is
part of a high-$p_T$ jet. But in those cases there are background processes,
which should be considered also and this requires more detailed study.  

\subsection{Neutral vector boson production}

In Eq.\ (\ref{h1h1bar}) 
we have given the contribution to the cross section that depends on 
the sum of the two transverse spin angles with respect to the lepton pair
production plane, i.e.\ $\cos(\phi_{S_1}^\ell+\phi_{S_2}^\ell)$. This means
that if one integrates over the lepton pair orientation, then this azimuthal
dependence will average to zero. At order $1/Q^2$ both quarks and gluons 
can contribute to $A_{TT}$ via a term in
the cross section which does not depend on the lepton scattering plane
\beq
A_{TT} \propto {\cos(\phi_{S_1}^\ell-\phi_{S_2}^\ell)}\, 
{\frac{M_1M_2}{Q^2} \, g_T \, \overline g_T},
\label{gTgT}
\eeq
but this is expected to be negligible at $Q^2=M_Z^2$. 
Moreover, it is not at all clear
that such a factorized description of the asymmetry holds at this level of
next-to-next-to-leading twist, since it is well known that for the unpolarized
case this order [${\cal O} (1/Q^4)$] does not factorize. 

On the other hand, if one were to observe the transverse momentum of the
lepton pair compared to the protons, there will be a double
transverse spin asymmetry as a function of this transverse momentum
$\bm{q}_T$,  
which appears at the leading order in $1/Q$. It will involve one more
angle ($\phi_h^\ell$), but even if one would integrate over this
angle (keeping only the magnitude of $\bm{q}_T$) and over the lepton pair 
orientation ($\phi^\ell$), then there will remain 
an azimuthal dependence in the cross section
that depends on the orientations of the two transverse polarization vectors
only.  

To make this explicit we will look at Eq.\ (A1) 
of Ref.\ \cite{Boer-99}, which gives the cross section for 
the polarized Drell-Yan process $p^\uparrow
\, p^\uparrow \to \gamma \, (Z) \, X$ in the
formalism of Ralston and Soper \cite{Ralst-S-79,Tangerman-Mulders-95a} 
using transverse momentum 
dependent distribution functions. From the
expressions for the production of the $Z$ boson it is easy to obtain the
expressions for the production of the $W$ boson. We will not repeat all the
details of the calculation of the cross section expressions, rather we will
focus on the expression 
as given in the Appendix in Ref.\ \cite{Boer-99}. For the
sake of argument it is unimportant to include contributions from the 
(formally difficult) 
$T$-odd distribution functions, hence we neglect them, but they can be easily
included. This leaves     
\ba
\frac{d\sigma(h_1h_2\to \ell \bar\ell X)}
     {d\Omega dx_1 dx_2 d^2{\bm q_T^{}}}&=&
\frac{\alpha^2}{3Q^2}\;\sum_{a,\bar a} \;\Bigg\{ 
- K_1^a(y)\;|\bm S_{1T}^{}|\;
                          |\bm S_{2T}^{}|\;
        \sin(\phi_h^\ell-\phi_{S_1}^\ell)\;\sin(\phi_h^\ell-\phi_{S_2}^\ell)\;
             {\cal F}\left[\,\frac{\bm p_T^{}\!\cdot \!
                      \bm k_T^{}}{M_1M_2}\,
                    g_{1T} \overline g_{1T}\right]\nonumber\\ 
&& - K_1^a(y)\;|\bm S_{1T}^{}|\; |\bm S_{2T}^{}|\; 
\cos(2\phi_h^\ell-\phi_{S_1}^\ell-\phi_{S_2}^\ell)\;
             {\cal F}\left[\,\frac{\bm{\hat h}\!\cdot \!\bm p_T^{}\,
                    \,\bm{\hat h}\!\cdot \!\bm k_T^{}}{M_1M_2}\,
                    g_{1T} \overline g_{1T}\right] + \ldots \Bigg\},
\label{xs}
\ea
where 
\beq
K_1(y)= \left(\frac{1}{2} -y+y^2\right) \;
\left[ e_a^2+ 2 g_V^l e_a g_V^a \chi_1 + c_1^l c_1^a \chi_2
\right] 
- \frac{1-2y}{2} \; \left[ 2 g_A^l e_a g_A^a \chi_1 + c_3^l c_3^a \chi_2 
\right],
\eeq
which contain the combinations of the couplings
\begin{eqnarray} 
c_1^j &=&\left(g_V^j{}^2 + g_A^j{}^2 \right),
\\
c_3^j &=&2 g_V^j g_A^j.
\end{eqnarray}
The $Z$-boson propagator factors are given by
\ba
\chi_1 &=& \frac{1}{\sin^2 (2 \theta_W)} \, \frac{Q^2
(Q^2-M_Z^2)}{(Q^2-M_Z^2)^2 + \Gamma_Z^2 M_Z^2},\\
\chi_2 &=& \frac{1}{\sin^2 (2 \theta_W)} \, \frac{Q^2}{Q^2-M_Z^2} \chi_1,
\ea
and $g_V$ and $g_A$ are the vector and axial-vector couplings to the $Z$
boson.
We have summed over the polarization of the outgoing leptons.
Furthermore, we use the convolution notation 
(Ralston and Soper \cite{Ralst-S-79} use $I[...]$) 
\begin{equation} 
{\cal F}\left[f\overline f\, \right]\equiv \;
\int d^2\bm p_T^{}\; d^2\bm k_T^{}\;
\delta^2 (\bm p_T^{}+\bm k_T^{}-\bm 
q_T^{})  f^a(x_{1},\bm{p}_T^2) 
\overline f{}^a(x_{2},\bm{k}_T^2),
\label{conv}
\end{equation}
where $a$ is the flavor index.

The function $g_{1T}$ is the function $h^{LT}$ of Ralston and Soper 
\cite{Ralst-S-79}
and has as interpretation the distribution of {\em longitudinally\/} polarized
quarks ($\gamma^+ \gamma_5$ projection) inside a transversely polarized 
hadron. It enters into the 
calculation compared to the unpolarized distribution function as follows:
\begin{eqnarray}
\Phi(x_1,\bm{p}_T) &=& 
\frac{M_1}{2P_1^+}\,\Biggl\{
f_1(x_1 ,\bm p_T^2)\, \frac{\slsh{\! P_1}}{M_1} 
- \frac{(\bpt\cdot\bm{S}_{1T})}{M_1}\,g_{1T}(x_1 ,\bm p_T^2)\, 
\frac{\slsh{\! P_1} \gamma_5}{M_1} + \ldots \Biggl\}.
\end{eqnarray}
The details of the momenta are: the 
momenta of the quarks which annihilate into the photon with momentum $q$, 
are predominantly along the direction of the parent
hadrons. One hadron momentum ($P_1$) is chosen to be along the 
lightlike 
direction given by the vector $n_+$ (apart from mass corrections). 
The second hadron with momentum $P_2$ 
is predominantly in the $n_-$ direction which satisfies $n_+ \cdot n_- = 1$, 
such that $P_1 \cdot P_2 = {\cal O} (Q^2)$. We decompose the momenta in $+, -$
and transverse components, defined through $p^\pm=p\cdot n_\mp$, where we
note that [cf.\ Eqs.\ (\ref{tvector})--(\ref{hvector})]
\begin{eqnarray}
n_+^\mu & = & 
\frac{1}{\sqrt{2}} \left[ \hat t^\mu + \hat z^\mu
-\,\frac{Q_T^{}}{Q} \hat h^\mu \right], \label{nplusc}
\\
n_-^\mu & = & 
\frac{1}{\sqrt{2}} \left[ \hat t^\mu - \hat z^\mu
-\,\frac{Q_T^{}}{Q}\,\hat h^\mu \right].  \label{nplusc2}
\end{eqnarray} 
The four-momentum conservation delta function at the vector boson vertex is 
written as (neglecting $1/Q^2$ contributions)
\begin{equation} 
\delta^4(q-k-p)=\delta(q^+-p^+)\, \delta(q^--k^-)\, \delta^2(
\bm{p}_T^{}+
\bm{k}_T^{}-\bm{q}_T^{}),
\label{deltafn0}
\end{equation}
and allows for integration over $p^-$ and $k^+$. 
However, the transverse momentum integrations cannot be separated, unless one 
integrates
over the transverse momentum of the vector boson or --equivalently-- of the 
lepton pair.

The Drell-Yan cross section 
is obtained by contracting the lepton tensor with the hadron tensor. At the 
tree level we find, for the hadron tensor, 
\beq
{\cal W}^{\mu\nu}=\frac{1}{3} \int d^2\bm{p}_T^{} d^2 
\bm{k}_T^{}\, \delta^2(\bm{p}_T^{}+
\bm{k}_T^{}-\bm{q}_T^{})\, \left.
\text{Tr}\left( \Phi (x_1,\bpt) \, 
V_1^\mu \, \overline \Phi  
(x_2,\bkt) \, V_2^\nu \right) \right|_{p^+, \, k^-}
+ \left(\begin{array}{c} 
q\leftrightarrow -q \\ \mu \leftrightarrow \nu
\end{array} \right).
\label{hadrontensor}
\eeq
The vertices $V_i^\mu$ can be either the photon vertex 
$V^\mu = e \gamma^\mu$ or the
$Z$-boson vertex $V^\mu = g_V \gamma^\mu + g_A \gamma_5 \gamma^\mu$.

The above given azimuthal dependence in the cross section, Eq.\ (\ref{xs}), 
means that if one
observes the transverse momentum of the $\gamma$ or $Z$ boson, one can
consider the cross section differential in the magnitude of the transverse
momentum only and integrate over the orientations of the leptons and of
$\bm q_T$ itself. This results in the following double transverse spin 
asymmetry:
\beq
A_{TT}(Q_T) 
= \frac{d\sigma\left[p^{\uparrow} \, {p^{\uparrow}} \to \gamma \, (Z) \,
X \right]  
\, {-} \, d\sigma\left[p^{\uparrow} \, {p^{\downarrow}} \to \gamma \, (Z) \,
X\right] }{
d\sigma\left[p^{\uparrow} \, {p^{\uparrow}} \to \gamma \, (Z) \, X\right] 
\, {+} \, d\sigma\left[p^{\uparrow} \, {p^{\downarrow}} \to \gamma \, (Z) \,
X\right]} =
- \frac{\sum_{a,\bar a}\;K_1^a(y)\; {\cal F}\left[\,
                    \bm p_T^{}\!\cdot \!
                      \bm k_T^{}\, 
g_{1T} \overline g_{1T}\right] }{2{M_1M_2}\, \sum_{a,\bar a}\; K_1^a(y) \; 
{\cal F}\left[ f_1 \; \overline
f_1 \right] },
\label{attwqt}
\eeq
where $Q_T^2 \equiv - q_T^2 \equiv \bm{q}_T^2 \ll Q^2$.
This can be seen from the following considerations. The angular dependence 
$\sin(\phi_h^\ell-\phi_{S_1}^\ell)\;\sin(\phi_h^\ell-\phi_{S_2}^\ell)$
can be rewritten after integration
over the angle $\phi_h^\ell$ as 
$\cos(\phi_{S_1}^\ell-\phi_{S_2}^\ell)/2=\cos(\phi_{S_2}^{S_1})/2$. 
Since this does not depend on the orientation of $\ell$ itself, one can 
integrate over it also. The angular dependence
$\cos(2\phi_h^\ell-\phi_{S_1}^\ell-\phi_{S_2}^\ell)$ averages out. 

If on the other hand one only observes the angle of the transverse momentum
and averages over its magnitude, one can also obtain a nonvanishing asymmetry
(which can still be differentiated from the transversity asymmetry). The whole
point is to prevent the averaging: $\int d^2 \bm{q}_T^{} \, 
{\cal F}\left[\,\bm p_T^{}\!\cdot \!\bm k_T^{}\, 
g_{1T} \overline g_{1T}\right] = 0 $. As said before, an incomplete averaging 
due to imposed experimental cuts might also result in a nonvanishing
asymmetry, cf.\ for instance Ref.\ \cite{Mirkes}. 

Before we continue we would like to point out that unlike for $h_1$
there is a leading twist gluon analogue of the function $g_{1T}$. The 
function arises with $i \epsilon_T^{\alpha \beta}$ in
the correlation function $\Phi^{\alpha \beta} \propto \amp{PS|F^{+\alpha}
F^{+\beta}|PS}$, which means that it is a
$\Delta g$ type of function with transverse momentum dependence. But since 
the transition $g\, g\to \gamma(Z)$ is only possible via quarks, we 
implicitly include the gluon in the sum over flavors.

\subsection{Charged vector boson production}

In order to arrive at the expressions for the cross sections
of the charged current process (cf.\ Ref.\ \cite{Boer-J-M-99}), one can 
take $e_a=0$ and replace 
\beq
\chi_{2}^Z \to \chi_{2}^W = \left( \frac{1}{8 \sin^2 \theta_W}\right)^2  \, 
\frac{Q^4}{(Q^2-M_W^2)^2 + \Gamma_W^2 M_W^2} \;, 
\eeq
in the above given coupling $K_1^a$. In addition, one 
replaces $c_1=\pm c_3 = 1$, depending on the chirality of the 
quark or lepton, since $c_1= (g_R^2+ g_L^2)/2$ and 
$c_3= (g_L^2- g_R^2)/2$. Hence, for a left-handed
quark one finds $c_1^a= c_3^a=1$ and for a right-handed
quark one finds $c_1^a= -c_3^a=1$; similarly for the leptons. We also
note that a left- or right-handed quark or lepton has helicity 
$\lambda_{q/e}=\mp 1$. This results in 
\ba
K_1^{ab}(y)&=& 4 \, \chi_{2}^W |V_{ab}|^2
\left(\frac{1}{2}-y+y^2 - \lambda_q \lambda_e \frac{1-2y}{2} 
\right) \nn\\
& = & 4 \, \chi_{2}^W |V_{ab}|^2 \, \times \Biggl\{ 
\begin{array}{cl} y^2 & \quad \text{for equal quark and lepton chiralities},
\\ (1-y)^2 & \quad \text{for opposite quark and lepton chiralities}, 
\end{array} 
\label{K12-charged}
\ea
where $a,b$ are the incoming quark and antiquark flavor indices, respectively, 
and $V_{ab}$ stands for the appropriate Cabibbo-Kobayashi-Maskawa (CKM) 
matrix element. We illustrate the
above by assuming that only $u$ and $d$ quark distribution functions
contribute. This leaves two elementary subprocesses:  
$u \bar d \to W^+ \to e^+ \nu$ ($u$ and $\nu$ have equal chiralities) 
and $d \bar u \to W^- \to e^- \bar\nu$ ($d$ and $e^-$ have equal 
chiralities) for which one finds the couplings $K_1^{u \bar d} = K_1^{\bar u d}
= 4 \, \chi_2^W |V_{ud}|^2 y^2$. For the cross section one has to take into 
account that in the sum over final state polarizations there is now only one
state that contributes, but for the asymmetry this is not relevant. 

In case of $p^\uparrow  \, p^\uparrow \to W \, X$ and subsequent leptonic
decay of the $W$, we encounter the problem 
that the produced neutrino will prevent a determination of the transverse
momentum of the $W$ boson. Hence, in the case of a produced neutrino 
one cannot define a lepton scattering plane 
(one does not observe $l'$), hence
azimuthal angles cannot be defined compared to the $W$ boson direction. This
holds unless one can reconstruct the direction 
of the neutrino by the momentum imbalance \cite{Ahmed}. 

Another possible way of observing the transverse momentum of the $W$ boson is 
looking at the $W$ decay into 2 jets. The expressions for lepton pair 
production stay
essentially the same for the 2 jets case after the obvious replacement of the
coupling constants. By measuring the direction of the 2 jets, the
transverse momentum of the $W$ can be determined, but one problem is that it
has $\gamma^*/Z \to 2\, \text{jets}$ as a background. Separation of
$\gamma^*/Z$ and $W$ might only be possible with a very high transverse
momentum cut \cite{Saito}, but then the given expressions are not
applicable anymore. Another problem is that it also receives contributions from
quark-quark scattering next to the quark-antiquark scattering, but such 
contributions have rather large color factor suppression
\cite{Ji,Jaffe-Saito}. In any case, the contribution coming 
from the transversity functions to
the $2\, \text{jets}$ asymmetry can always be eliminated by averaging 
over the orientation of the 2 jets, as explained above. 

Here we will focus only on lepton pair production and assume that in case of 
$W$ production the direction of the neutrino can be reconstructed to obtain 
the transverse momentum of the $W$. 

\subsection{Gaussian transverse momentum dependence}

In order to obtain an estimate of the above asymmetry, we will consider a
Gaussian transverse momentum dependence of the functions. 
Instead of using Gaussians, another way of obtaining an estimate of the 
asymmetry would be to use the spectator
model for the function $g_{1T}(x,\bm p_T^2)$ \cite{Joao}.
But for simplicity we will assume a Gaussian
transverse momentum dependence, e.g.\ 
\beq
g_{1T}(x,\bm p_T^2) = g_{1T}(x) \; \frac{R^2}{\pi} \; \exp\left(-R^2 
\bm p_T^2\right) \equiv g_{1T}(x) \; {\cal G}(\bm{p}_T^2).
\label{Gauss}
\eeq  

We would like to relate the function $g_{1T}(x)$ to a well-known function in
order to be able to make some predictions in the end. This can be achieved
by using $\bm{p}_T^2$ weighted functions
\beq
f^{(1)}(x)= \int d^2 \bpt \, \frac{\bm p_T^2}{2M^2} \, 
f(x, \bm p_T^2).
\eeq
For a Gaussian transverse momentum dependence we find that 
$g_{1T}^{(1)}= g_{1T}(x)/(2M^2R^2)$. In the Wandzura-Wilczek approximation 
the function $g_{1T}^{(1)}$ is a 
well-known quantity: it equals (upon neglecting quark
masses) $x \, g_T^{WW}(x)$, where $g_T^{WW}(x)= g_1 + g_2^{WW} $ is the
Wandzura-Wilczek part of the function $g_T$ (see also Ref.\ \cite{Boer-99b} 
for a discussion on this topic). This can be shown by using the equations of 
motion. The function $g_T$ has been studied 
by SLAC and the Spin Muon Collaboration (SMC) \cite{SLAC} and the data are 
(still) consistent with 
$g_T= g_T^{WW}$. Also, the data show that $g_2$ is small 
compared to $g_1$, therefore, up to a few percent one can take 
$x \, g_T^{WW}(x) \approx x \, g_1(x)$. Thus, we find 
$g_{1T}(x)= g_{1T}^{(1)}\, 2M^2R^2 \approx x g_1(x) \, 2M^2R^2$. 

Furthermore, will assume
that the Gaussians are the same for $f_1$ and $g_{1T}$ and for both protons,
i.e., we take $R_1=R_2=R$ and $M_1=M_2=M$. 
In this way we find for example 
\beq
{\cal F}\left[\,\bm p_T^{}\!\cdot \!
                      \bm k_T^{}\, g_{1T} \overline g_{1T}\right]
\approx 
-\frac{M^4 R^4}{\pi}\left( 1 -\frac{Q_T^2 R^2}{2} \right) \exp\left(-
\frac{Q_T^2 R^2}{2}\right) \; x_1 g_{1}(x_1) \; x_2 \overline g{}_{1}^{}(x_2).
\label{Gaussconv}
\eeq
This results in the following tree level double transverse spin 
asymmetry at $Q_T=0$
\beq
A_{TT}(Q_T=0) 
= M^2 R^2 \; \frac{\sum_{a,\bar a}\;K_1^a(y)\; 
x_1 g_1^a(x_1) \; x_2 \overline g{}_{1}^a(x_2)}{\sum_{a,\bar a}\;
K_1^a(y)\; f_1^a(x_1) \; \overline f{}_1^a(x_2)}.
\label{treeattwqt}
\eeq

We will also encounter the Fourier transforms of these functions.  
The function $\tilde{f}$ will denote the Fourier transform of $f$, and 
since we use the notation $f(x)= \int d^2 \bm{p}_T^{} \,
f(x,\bm{p}_T^{})$, we see that $f(x)= \tilde{f}(x,b=0)$. 
Taking the Fourier transform of Eq.\ (\ref{Gauss}) yields
\beq
\tilde{g}{}_{1T}^{}(x,b^2) 
= g_{1T}(x) \; \exp\left(-\frac{b^2}{4 R^2} \right).
\label{GaussFT}
\eeq  

\subsection{Beyond the range of intrinsic transverse momentum}

Monte Carlo studies including soft gluon resummation \cite{Balazs} show that 
the largest contribution to
the unpolarized cross section arises when the transverse momentum of the $W$ is
several GeV. This transverse momentum is too high to trust tree 
level expressions which involve only intrinsic transverse momenta. 
In order to go beyond this region, we will also include the Sudakov factor 
arising from resummed perturbative corrections to the transverse momentum
distribution.  

Resummation of soft gluons into
Sudakov form factors \cite{Col-89} results in a replacement in Eqs.\
(\ref{conv}) and (\ref{hadrontensor}) of 
\beq
\delta^2(\bm{p}_T^{}+\bm{k}_T^{}-\bm{q}_T^{})\to \int 
\frac{d^2 \bm{b}}{(2\pi)^2} \, e^{-i \bm{b} \cdot
(\bm{p}_T^{}+\bm{k}_T^{}-\bm{q}_T^{})} \, e^{-S(b)}, 
\label{sudakovreplacement}
\eeq 
where $e^{-S(b)}$ is the Sudakov form factor and $b^2=\bm{b}^2$. 
This has been shown in Refs.\ \cite{CSS83,CSS85} for the leading twist and is
discussed for the present context in more detail in Ref.\
\cite{Boer-Mulders-00}. 
The Sudakov form factor is found to be 
\beq
S(b,Q)=\int_{b_0^2/b^2}^{Q^2} \frac{d \mu^2}{\mu^2} \left[ 
A(\alpha_s (\mu)) \, \ln \frac{Q^2}{{\mu}^2} + 
B (\alpha_s (\mu)) \right].
\eeq
One can expand the functions $A$ and $B$ in $\alpha_s$ and the first few 
coefficients are known for unpolarized scattering 
\cite{Davies-Stirling} and for longitudinally polarized
scattering \cite{Weber}. The latter result is needed here since the function
$g_{1T}$ is a distribution of longitudinally polarized quarks; the asymmetry 
on the parton level is $a_{LL}$. In order to
obtain a first estimate of the effect of including the Sudakov factor we will 
take into account only the 
first term in the expansion of $A$: $A^{(1)} =  C_F/\pi$. This leads to the
expression \cite{Frixione}
\beq
S(b,Q)=-\frac{16}{33-2n_f} \left[ \log\left(\frac{b^2 Q^2}{b_0^2}\right)+
\log\left(\frac{Q^2}{\Lambda^2}\right)\; \log\left[1- \frac{\log\left(b^2 
Q^2/b_0^2\right)}{\log\left(Q^2/\Lambda^2\right)} \right]\right],
\eeq
with $b_0=2\exp(-\gamma_E) \approx 1.123$. 
We will take for the number of flavors $n_f =5$ and also $\Lambda_{QCD}=200 \,
\text{MeV}$. 

The replacement in Eq.\ (\ref{conv}) leads to (suppressing the flavor index) 
\ba 
{\cal F}\left[f\overline f\, \right]& \equiv & \; \int 
\frac{d^2 \bm{b}}{(2\pi)^2} \, e^{i \bm{b} \cdot \bm{q}_T^{}} 
\, e^{-S(b)}\, \tilde{f}(x_{1},b) \, 
\tilde{\overline f}(x_{2},b)\nn \\[2 mm]
&=& \frac{1}{2 \pi} \int_0^\infty db \, b \, J_0(b Q_T)\, e^{-S(b)}
\tilde{f}(x_{1},b) \, \tilde{\overline f}(x_{2},b).
\label{conv2}
\ea
The functions also have a renormalization and 
factorization scale dependence, which we will choose to be equal 
$\mu_R=\mu_F=\mu$. Hence, we have, for instance,
\beq
f(x;\mu)= \int d^2 \bm{p}_T^{} \, f(x,p_T;\mu) \equiv \tilde{f}(x,b=0;\mu),
\eeq
where also the boundary of the integration gives a $\mu$ dependence. 
In Eq.\ (\ref{conv2}) one usually takes $\tilde{f}(x_{1},b;\mu = b_0/b)$
\cite{CSS85}. 

Of course, if one includes the effects of perturbative corrections, one should
also include higher order corrections to the hard part not coming from soft
gluons. But since the formalism, which means the factorized formula, is 
valid only for $\bm{q}_T^2 \ll Q^2$, we include only the effects of soft
gluons, which should allow us to extend the range of applicability from the 
region of intrinsic transverse
momentum to the region of moderate $\bm{q}_T^{}$ values. 
An equivalent way of saying this is that one can perform
a collinear expansion of the hard scattering part 
$H^{\mu\nu}(x_1,x_2,\bm{p}_T^{},\bm{k}_T^{},\bm{q}_T^{},Q) 
\approx H^{\mu\nu}(x_1,x_2,Q)$, such
that perturbative corrections to the hard part do not affect the transverse 
momentum structure of the tree level result and
hence the transverse momentum weight in the asymmetry will be the same also
beyond the tree level. Here we will consider $H^{\mu\nu}(x_1,x_2,Q)$ to lowest
order in $\alpha_s$, therefore, only logarithmic $Q^2$ corrections to the
results presented below are expected. 

The numerator in Eq.\ (\ref{attwqt}) 
cannot be treated exactly like the denominator, so let us focus next on
\ba 
{\cal F}\left[\, \bm p_T^{} \cdot \bm
k_T^{} \; f\overline f\, \right]& \equiv & \; \int 
\frac{d^2 \bm{b}}{(2\pi)^2} \, e^{i \bm{b} \cdot \bm{q}_T^{}} 
\, e^{-S(b)}\, \int d^2\bm p_T^{}\; d^2\bm k_T^{}\; \bm p_T^{} \cdot \bm
k_T^{} \; e^{-i \bm{b} \cdot (\bm{p}_T^{}+\bm{k}_T^{})}\; f(x_{1},\bm{p}_T^2) 
\overline f(x_{2},\bm{k}_T^2).
\label{bconv}
\ea
As mentioned before, we assume that the distribution functions are
Gaussians (as a function of transverse momentum), all of equal width: 
$f(x_{1},\bm{p}_T^2)= f(x_{1}) \; {\cal G}(\bm{p}_T^2)$ and 
$\overline f(x_{2},\bm{k}_T^2)= \overline f(x_{2}) \; 
{\cal G}(\bm{k}_T^2)$. One can then change variables 
to $\bm u=(\bm{p}_T^{}+\bm{k}_T^{})/\sqrt{2}$ 
and $\bm v= (\bm{p}_T^{}-\bm{k}_T^{})/\sqrt{2}$ and compute 
\ba   
\int d^2\bm p_T^{}\; d^2\bm k_T^{}\; \bm p_T^{} \cdot \bm
k_T^{} \; e^{-i \bm{b} \cdot (\bm{p}_T^{}+\bm{k}_T^{})}\; {\cal
G}(\bm{p}_T^2) \;  
{\cal G}(\bm{k}_T^2) 
&=& -\frac{b^2}{4 R^4} \; \exp\left(-\frac{b^2}{2 R^2}\right),
\ea 
which after application to Eq.\ (\ref{bconv}) yields 
(see also Eq.\ (\ref{GaussFT}))
\beq
{\cal F}\left[\, \bm p_T^{} \cdot \bm
k_T^{} \; f\overline f\, \right] = - \int 
\frac{d^2 \bm{b}}{(2\pi)^2} \, e^{i \bm{b} \cdot \bm{q}_T^{}} \,
\frac{b^2}{4 R^4} \; 
e^{-S(b)}\tilde{f}(x_{1},b) \, 
\tilde{\overline f}(x_{2},b), 
\eeq
which can be compared with Eq.\ (\ref{Gaussconv}).  
Both equations 
fulfill the property that the expression should vanish after $\bm q_T^{}$ 
integration. One infers that 
the numerator of $A_{TT}(Q_T^{})$ and hence $A_{TT}(Q_T^{})$ itself 
oscillate, but in general $\int d Q_T^{} \,A_{TT}(Q_T^{}) \neq 0$, because of
the $Q_T^{}$ dependence of the denominator. 
 
For the asymmetry, Eq.\ (\ref{attwqt}), we then obtain
\ba
A_{TT}(Q_T) & = & \frac{1}{8 M^2 R^4}\,
\frac{\sum_{a,\bar a}\;K_1^a(y)\; 
\int_0^\infty db \, b^3 \, J_0(b Q_T)\, e^{-S(b)}\, 
\tilde{g}{}_{1T}^a(x_{1},b) \, 
\tilde{\overline g}{}_{1T}^a(x_{2},b)}{\sum_{a,\bar a}\; 
K_1^a(y) \; 
\int_0^\infty db \, b \, J_0(b Q_T)\, e^{-S(b)}
\, \tilde{f}{}_1^a(x_{1},b) \, 
\tilde{\overline f}{}_1^a(x_{2},b)}\nn \\[2 mm]
& \approx & \frac{M^2}{2} \, \frac{\sum_{a,\bar a}\;K_1^a(y)\;  
x_1 g_1^a(x_1) \; x_2 \overline g{}_{1}^a(x_{2})}{\sum_{a,\bar a}\; 
K_1^a(y) \; f_1^a(x_1) \; 
\, \overline f{}_1^a(x_{2})}\, \frac{\int_0^\infty db \,b^3 \, J_0(b Q_T)\, 
\exp\left[{-S(b)}\, -\frac{1}{2} \,b^2/R^2 \right]}{\int_0^\infty db \, 
b \, J_0(b Q_T)\, \exp\left[{-S(b)} \, -\frac{1}{2} \,b^2/R^2 \right]},
\ea
where the approximation arises from taking $g_{1T}(x)= g_{1T}^{(1)}\, 2M^2R^2 
\approx x g_1(x) \, 2M^2R^2$. 

In order to extend the above equation to the nonperturbative region of
large values of $b$, one usually introduces 
$b_*=b/\sqrt{1+b^2/b_{\max}^2}$ and an additional term 
$\exp\left[{-S_{NP}(b)}\right]$, needed to describe the low $\bm{q}_T$ region 
properly. 
In part, $S_{NP}(b)$ is introduced to take care of the 
smearing due to
the intrinsic transverse momentum, therefore, taking into account the term 
$\exp\left(-\frac{1}{2} \, b^2/R^2 \right)$ in addition will just produce a 
change in the coefficient of the $b^2$ term in $S_{NP}(b)$. 
To keep the unpolarized 
cross section unaffected, we will therefore introduce
as nonperturbative term $\exp\left[{-S_{NP}(b)}
\, + \frac{1}{2}\,b^2/R^2 \right]$ and study 
the following final expression for the asymmetry:
\beq
A_{TT}(Q_T) = \frac{1}{2} \, \frac{\sum_{a,\bar a}\;K_1^a(y)\;  
x_1 g_1^a(x_1) \; x_2 \overline g{}_{1}^a(x_{2})}
{\sum_{a,\bar a}\; 
K_1^a(y) \; f_1^a(x_1) \; 
\, \overline f{}_1^a(x_{2})}
\;{\cal A}(Q_T),
\eeq
where we define
\beq
{\cal A}(Q_T) \equiv M^2 \, \frac{
\int_0^\infty db \, b^3 \, J_0(b Q_T)\, \exp\left[{-S(b_*)}\, 
{-S_{NP}(b)}\right]}{\int_0^\infty db \, b \, J_0(b Q_T)\, 
\exp\left[{-S(b_*)}\, {-S_{NP}(b)} \right]}.
\eeq
The denominator is
then the conventional unpolarized expression. Also, we note that ${\cal
A}(Q_T)$ is dimensionless and, for simplicity, we will take 
$M= 1 \, \text{GeV}$. 

The above approach of including Sudakov factors in the tree level azimuthal
asymmetry expressions can also be applied to expressions derived in
\cite{Boer,BoerZ}
for electron-positron annihilation and in \cite{Boer-J-M-99} for lepton-hadron
scattering. 

\subsection{Estimating the asymmetry}

For the case of $W$ production the asymmetry becomes 
\beq
A_{TT}^W(Q_T) = \frac{1}{2} \,\frac{\sum_{a,\bar a;b, \bar b}\;|V_{ab}|^2 \;  
x_1 g_1^a(x_1) \; x_2 \overline g{}_{1}^b(x_{2})}
{\sum_{a,\bar a;b, \bar b}\; 
|V_{ab}|^2 \; f_1^a(x_1) \; 
\, \overline f{}_1^b(x_{2})}
\;{\cal A}(Q_T).
\label{attwqtfinal}
\eeq
If only the $u$ and $d$ quarks contribute, then also the CKM matrix 
elements drop out of the ratio. 

For the nonperturbative Sudakov factor we use the parameterization of 
Ladinsky-Yuan, Ref.\ \cite{Ladinsky-Yuan}, which was fitted to relevant 
Fermilab data, 
\beq
S_{NP}(b) = g_1b^2 + g_1 g_3 b \ln(100 x_1 x_2) + g_2 b^2
\ln\left(\frac{Q}{2Q_0}\right),
\eeq
with $g_1=0.11\,\text{GeV}^{2}, g_2=0.58\,\text{GeV}^{2}, 
g_3=-1.5\,\text{GeV}^{-1}, Q_0=1.6 \,\text{GeV}$ and 
$b_{\max}= 0.5 \,\text{GeV}^{-1}$. We will take $ x_1 x_2 =10^{-2}$, which is 
justified below. 
This leads to $S_{NP}(b) = 1.98 \, b^2$ at $Q =
80 \, \text{GeV}$ and $S_{NP}(b) = 0.77 \, b^2$ at $Q = 10 \, \text{GeV}$.

\begin{figure}[htb]
\begin{center}
\leavevmode \epsfxsize=8cm \epsfbox{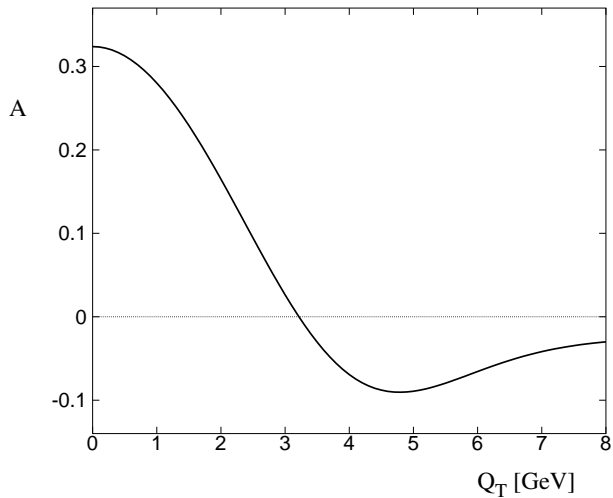}
\caption{\label{Attw80}The asymmetry factor ${\cal A}(Q_T)$ at $Q =
80 \, \text{GeV}$.}
\end{center}
\vspace{-2 mm}
\end{figure}

The result for the asymmetry factor ${\cal A}(Q_T)$ at $Q =
80 \, \text{GeV}$ is given in Fig.\ \ref{Attw80}. 
It is plotted up to 
$Q_T =Q/10$, since beyond that $Q_T$ range the magnitude only slowly 
decreases and also the approximation $Q_T^2\ll Q^2$ is expected to become 
less valid.
The asymmetry factor ${{\cal A}(Q_T)}$ at $Q_T=0$ is seen to be around 
$0.32$ and at $Q_T$ values of
a few GeV --relevant for the majority of produced $W$ bosons-- 
the asymmetry factor has a sign change and consequently a smaller magnitude. 
On top of that the
asymmetry is proportional to $|x_1 g_1(x_1) \, x_2 \overline g_1(x_2)| \leq 
x_1 f_1(x_1) \, x_2 \overline f_1(x_2)$. Therefore, the total asymmetry as a
function of $Q_T$ is expected to be below the percent level,
if one assumes that on average $x_1 = 0.4$ and $x_2 = 0.07$ for $W$ 
production at RHIC \cite{Saito}. 
Since we have implicitly included the gluons in the sum over flavors, the
latter argument is not valid if $\Delta g$ turns out to be extremely large at 
small $x$. Of course this will have even more serious implications for e.g.\ 
$A_{LL}$ in jet production, especially at low transverse momenta. We will just 
view this unlikely option as a proviso. 

In Ref.\ \cite{Martin} it is demonstrated that the transversity double
spin asymmetry, which is a $\bm{q}_T^{}$ integrated asymmetry, 
is expected to be at most on the order of a few 
percent, which matches the level of sensitivity of RHIC. The present asymmetry
is still a function of $Q_T$, requiring even more statistics. This will make
the asymmetry $A_{TT}^{W (Z)}(Q_T)$ invisible at RHIC. Moreover, 
since it oscillates as a function of $Q_T$, one expects that the asymmetry 
partly integrated over $Q_T$, will not lead to any significant result  
either.

A few remarks about the dependence of the result on the nonperturbative 
parameters. The asymmetry factor is seen to decrease with increasing 
Gaussian smearing width. Taking a higher value of $b_{\max}$ and a lower value
of $x_1x_2$ both increase this width. The above --optimistic-- choices of 
$b_{\max}= 0.5 \,\text{GeV}^{-1}$ and $x_1 x_2 =10^{-2}$ are therefore
expected to overestimate the asymmetry factor somewhat.   

But at lower energies --where larger light cone momentum fractions can be
achieved-- this asymmetry for $\gamma^*$ production is still worth
investigating. 
\begin{figure}[htb]
\begin{center}
\leavevmode \epsfxsize=8cm \epsfbox{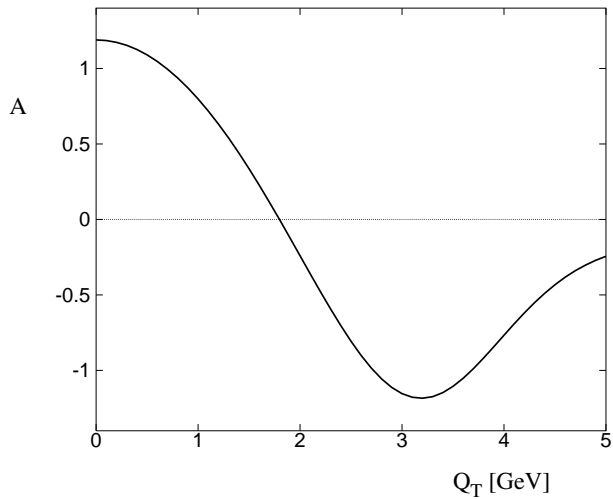}
\caption{\label{Attw10}The asymmetry factor ${\cal A}(Q_T)$ at $Q =
10 \, \text{GeV}$.}
\end{center}
\end{figure}
In Fig.\ \ref{Attw10} we have given the function 
${{\cal A}(Q_T)}$ at the scale $Q =
10 \, \text{GeV}$ and find that at low values of $Q_T$ it is around $1$. 
Measuring $A_{TT}^{\gamma^*}(Q_T)$ at larger values of $x_1$ and
$x_2$ might make this asymmetry observable. 

If one studies the values of ${{\cal A}(Q_T)}$ at $Q_T=0$ --where the
asymmetry is largest-- as a function of $Q$, one observes that inclusion 
of the Sudakov factor produces to good approximation a $1/Q^\alpha$ behavior,
with $\alpha \approx 0.6$. 
Even though this suppression is not very strong as a function of $Q$, 
one actually needs to compare the resulting asymmetry Eq.\ (\ref{attwqtfinal})
with the tree level asymmetry Eq.\ (\ref{treeattwqt}). 
This leads to a comparison of ${\cal A}(Q_T=0)$ and $2M^2 R^2$. In a tree level
analysis $R^2 = 1/\amp{\bm{p}_T^2}$ for a typical intrinsic transverse 
momentum squared value, i.e.\ $R^2 \approx 2 - 11 \, \text{GeV}^{-2}$,
corresponding to the range of $\amp{\bm{p}_T^2} \approx (300 - 700 \, 
\text{MeV})^2$. 
If we view $R^2 = 1$ as giving a lower bound for the tree level asymmetry 
factor, then ${\cal A}(Q_T=0, Q=80)$ is still an order of magnitude smaller.  

We conclude that the Sudakov factors produce a strong suppression compared to
the tree level result and the effect increases with energy. This 
will also have
consequences for similar types of transverse momentum dependent azimuthal
asymmetries appearing in for instance $e^+ e^-$ annihilation at the $Z$ mass
scale \cite{BoerZ,Efremov}, where the same strong suppression due to Sudakov
factors is expected. 

\subsection{Physics beyond the standard model}

It is now clear that the standard model (SM) 
mechanisms seem to produce negligible double transverse spin asymmetries in 
$W$ production ($A_{TT}^W$). In summary the reasons are the following: 
the transversity distribution $h_1$ does not contribute \cite{Bourrely-Soffer}.
At next-to-next-to-leading twist [${\cal O}(M_1 M_2/Q^2)$] the twist-3
distribution function $g_T$ (which is a helicity non-flip quark distribution)
can contribute and its gluon analogue $\Delta_T g$ as well. At $Q^2=M_W^2$
these contributions will be negligible.
Furthermore, one can also neglect contributions which are of 
higher order in the strong and/or weak coupling constants. 
In the case of perturbative QCD corrections double helicity flip will be 
accompanied by quark mass
terms and therefore will also be suppressed by at least two factors of $1/Q$. 
In the case of weak corrections one also expects 
negligible contributions, at least at RHIC energies, say around $Q^2=M_W^2$. 
Any leftover asymmetry from
incomplete averaging of the transverse momentum dependent asymmetry $A_{TT}^W 
(Q_T)$ we have
estimated to be negligible as well.  
So at RHIC energies $A_{TT}^W$ is expected to be negligible within the SM.  

It is now fair to address the question: 
if a significant asymmetry would nevertheless be found in the polarized 
proton-proton collisions at RHIC, can one really 
conclude something about physics beyond the SM? 
For instance, there could be scalar or tensor couplings of the $W$ to quarks
that can generate an asymmetry $A_{TT}^W$. 
If the scale of such new 
physics is $\Lambda \gg M_W$, then one might need to compare 
effects of order 
$M_1 M_2/Q^2$ with $Q^2/\Lambda^2$. For instance, for $Q^2=M_W^2$ and 
$\Lambda = 1 \, \text{TeV}$ the latter is a factor of 40 larger (although still
quite small). 
But the problem of comparing to higher twist contributions disappears
altogether if the new couplings violate symmetries. 

This means that on top of the fact that the various SM mechanisms produce 
negligible asymmetries,
one can also exploit the dependence on the orientation of the transverse spins
compared to the lepton production plane to cancel out specific contributions
exactly. We have seen that the transversity asymmetry $A_{TT}^{\gamma/Z}$ 
appears with an angular dependence ${\cos(\phi_{S_1}^\ell+\phi_{S_2}^\ell)}$,
whereas $A_{TT}(Q_T)$ and the $1/Q^2$ suppressed contribution from 
$g_T \overline g_T$ in Eq.\ (\ref{gTgT}) both appear with 
${\cos(\phi_{S_1}^\ell-\phi_{S_2}^\ell)}$. The latter does not
depend on the lepton scattering plane, because the asymmetries are not double
transverse spin asymmetries at the parton level. On the other hand, symmetry
violation asymmetries can produce other angular dependences than any SM
mechanism. 

There might be $T$-odd asymmetries, for
example the one of Ref.\ \cite{Rykov}, 
\beq
A_{TT}^{\perp} \propto {\sin(\phi_{S_1}^\ell+\phi_{S_2}^\ell)}, 
\eeq
which would arise due to $CP$-violating vector couplings of the 
quarks to the $W$, which is assumed not to be $V-A$ anymore, 
but some complex linear combination of $V$ and $A$. 
It can clearly be distinguished from possible 
initial state interaction effects, which are
$P$-even and only lead to asymmetries independent of the lepton scattering 
plane. 
To be more specific, 
if one assumes $T$-odd ($P$-even) distribution functions to be nonzero, 
then there will also be contributions proportional to [cf.\ Eq.\ (A1) of
\cite{Boer-99}] 
\ba
&& {\cos(\phi_{S_1}^\ell-\phi_{S_2}^\ell)} \, {{\cal F}}
\left[\, \bm p_T^{}\!\cdot \! \bm k_T^{}\, {
f_{1T}^\perp \overline f{}_{1T}^\perp } \right],  
\\[2 mm] 
&& {\sin(\phi_{S_1}^\ell-\phi_{S_2}^\ell)} \, {{\cal F}}
\left[\, \bm p_T^{}\!\cdot \! \bm k_T^{}\, {
f_{1T}^\perp \overline g_{1T} } \right]. 
\ea
The function ${f_{1T}^\perp}$ corresponds to the so-called 
{Sivers effect} \cite{s90}. Here one usually assumes that such a function
might arise due to initial state interactions and its contributions 
indeed do not depend on the lepton pair orientation as can be seen from 
the above two angular dependences. 

Therefore, these structures are distinguishable from the $T$-odd asymmetry $
A_{TT}^{\perp}$. However, it is important to note that $
A_{TT}^{\perp}$ can also effectively 
arise due to SM $CP$ violation, hence this contribution must first 
be calculated
before any conclusion about physics beyond the SM can be made. Also, 
this specific asymmetry will 
be accompanied by the product $h_1^a(x_1) \overline h{}_1^a(x_2)$, thus 
it will suffer from the same drawback as Eq.\ (\ref{h1h1}), namely that the
transversity function for the antiquarks is presumably smaller than for the
quarks, making this asymmetry hard (if not impossible) to detect at RHIC.
But it illustrates how symmetry violation can be used in principle 
to disentangle SM asymmetries from new physics asymmetries.  

\section{Conclusions}

We have investigated a helicity non-flip double transverse spin asymmetry in 
vector boson production in hadron-hadron scattering, which was first
considered by Ralston and Soper. It does not involve transversity 
functions and in principle also arises in $W$-boson production for which we
have presented the expressions. The asymmetry 
requires observing the transverse momentum of the vector boson, but it is not 
suppressed by explicit inverse powers of the large energy scale $Q$. 
However, as we have shown, inclusion of Sudakov factors suppresses the 
asymmetry at least by an order of magnitude compared to the tree level 
result. 
This suppression increases with energy approximately as a fractional 
power, numerically found to be $\alpha\approx 0.6$. 
Moreover, the asymmetry is shown to be
approximately proportional to $x_1 g_1(x_1) \, x_2 \overline g_1(x_2)$, 
which gives rise
to additional suppression at small values of the light cone momentum fractions.
This implies that the asymmetry is negligible for $Z$ and $W$ production
at RHIC and is mainly of interest at low energies (for $\gamma^*$ production). 
The strong suppression with respect to the tree level result will also have 
consequences for similar types of transverse momentum dependent azimuthal
asymmetries in for instance $e^+ e^-$ annihilation at the $Z$ mass
scale, where the same strong suppression due to Sudakov
factors is expected.

We have also noted that unlike the transversity and $CP$-violating double 
transverse spin asymmetries, the helicity
non-flip asymmetry $A_{TT}(Q_T)$ does not depend on the orientation of the 
transverse spin
vectors compared to the lepton pair production plane orientation. This feature
can be exploited to separate the different types of asymmetries. 

\acknowledgments 
I would like to thank Les Bland, Gerry Bunce, Bob Jaffe, Rainer Jakob, 
Piet Mulders, Naohito Saito and Werner Vogelsang for helpful comments and 
discussions. 
Furthermore, I thank RIKEN, Brookhaven National Laboratory and the U.S.\ 
Department of Energy (contract DE-AC02-98CH10886) for
providing the facilities essential for the completion of this work.

\end{document}